\documentstyle[multicol,epsf,aps,pre]{revtex}
\begin{document}
\title{Does Good Mutation Help You Live Longer?}
\author{W. Hwang, P. L. Krapivsky, and S. Redner}
\address{Center for BioDynamics, Center for Polymer Studies, and Department of Physics, 
Boston University, Boston, MA 02215}

\maketitle 
\begin{abstract} 
  We study the dynamics of an age-structured population in which the life
  expectancy of an offspring may be mutated with respect to that of its
  parent.  When advantageous mutation is favored, the average fitness of the
  population grows linearly with time $t$, while in the opposite case the
  average fitness is constant.  For no mutational bias, the average fitness
  grows as $t^{2/3}$.  The average age of the population remains finite in
  all cases and paradoxically is a {\em decreasing} function of the overall
  population fitness.

\end{abstract}

\begin{multicols}{2}
\narrowtext

In this letter, we investigate the role of mutation on the age distribution
and fitness of individuals within a simple age-structured population dynamics
model.  The basic feature of our model is that the life expectancy of an
offspring, which measures its fitness, may be mutated with respect to that of
its parent.  While age-structured population models have been studied
previously\cite{murray,charlesworth}, relatively little is known about the
role of mutation.  When the individual reproduction rate is the fitness
measure and population is regulated by externally imposed death, mutation
leads to predominance by the fittest species\cite{kessler}.  In a related
vein, it was recently shown that longevity is
heritable\cite{olivera,evolution} within the Penna bit-string model of
aging\cite{penna}.  In these studies, the role of positive mutations was
central.  Our focus is quite different, as we study the dynamics of the
fitness and age in a self-interacting population as a function of the
advantageous and deleterious mutation rates.

When advantageous mutation is favored, that is, the offspring life expectancy
(fitness) is greater than that of its parent, the fitness distribution of the
population approaches a Gaussian with average fitness growing linearly in
time and dispersion increasing as $t^{1/2}$.  Conversely, when deleterious
mutation is more likely, there is a $t^{-2/3}$ approach to a steady fitness
distribution.  In the absence of mutational bias, the fitness distribution
again approaches a Gaussian, with average fitness growing as $t^{2/3}$ and
width growing as $t^{1/2}$.  The average age of the population reaches a
steady value in all cases and, surprisingly, is a {\em decreasing} function
of the average fitness.  Therefore within our model, {\em a fitter population
  leads to a decreased individual lifetime.}

Our model is a simple population dynamics scenario which incorporates age
structure and mutation.  This dynamics is based on the logistic model, $\dot
N=b N-\gamma N^2$, in which a population with density $N(t)$ evolves both by
birth at rate $b$, and death at rate $\gamma N$, with steady-state solution
$N_\infty=b/\gamma$.  The crucial new element in our model is that the life
expectancy of each newborn may be mutated by $\pm\tau$ (with $|\tau|=1$
without loss of generality) with respect to that of its parent.  We also
assume a constant age-independent mortality rate and birth rate for each
individual.

Each of these features represent idealizations of reality; for example, it
would be more realistic to incorporate a mortality rate which is an
increasing function of age\cite{charlesworth,evolution,azbel}.  We shall
argue below that our choice of an age-independent mortality leads to behavior
which applies to systems with realistic mortality rates.  The nature of our
results also suggests that the details of the mutation-driven shift in
offspring life expectancy is not crucial.

Let $C_n(a,t)$ be the density of individuals with life expectancy $n\geq 1$
and age $a$ at time $t$.  According to our model, the rate equation for
$C_n(a,t)$ is
\begin{equation}
\label{cn}
\left({\partial \over\partial t}+{\partial \over\partial a}\right)C_n(a,t)
=-\left(\gamma N(t)+{1\over n}\right)C_n(a,t). 
\end{equation}
The derivative with respect to $a$ on the left hand side accounts for
aging\cite{murray,nisbet}.  On the right hand side, the loss term $\gamma
NC_n$ accounts for death by competition and is assumed to be independent of
an individual's age and fitness.  As discussed above, the mortality rate is
taken as age independent; the form $C_n/n$ guarantees that the life
expectancy equals $n$.

We account for the population of newborns as a boundary condition for
$C_n(a=0,t)$ \cite{murray}.  An individual produces offspring with the same
life expectancy at rate $b$, and, due to mutation, produces offspring whose
life expectancy is longer or shorter than its parent by $\pm 1$, with
respective rates $b_\pm$.  Defining $P_n(t)= \int_0^\infty da\,C_n(a,t)$ as
the density of individuals at time $t$ of any age whose life expectancy
equals $n$, then the boundary condition for $C_n(0,t)$ is
\begin{equation}
\label{boundary}
C_n(0,t)=bP_n(t)+b_+P_{n-1}(t)+b_-P_{n+1}(t). 
\end{equation}

To determine the asymptotic behavior of the age and fitness distributions, it
proves useful to first disregard the age structure and focus on fitness
alone.  From Eqs.~(\ref{cn})--(\ref{boundary}), the rate equations for
$P_n(t)$ for $n\geq 1$ are
\begin{equation}
\label{pn}
{d P_n\over dt}=\left(b-\gamma N-{1\over n}\right)P_n 
+b_+P_{n-1}+b_-P_{n+1},
\end{equation}
with $P_0=0$.  This describes a random-walk-like process in a one-dimensional
fitness space which is augmented by birth and death due to the first term on
the right-hand side.  Using $N(t)=\Sigma P_n(t)$, we find that the total
population density obeys a generalized logistic equation
\begin{equation}
\label{N}
{d N\over dt}=(B-\gamma N)N-\sum_{n=1}^\infty{P_n\over n} 
-b_-P_1,
\end{equation}
where $B\equiv b+b_++b_-$ is the total birth rate.

We now discuss the asymptotic behavior of these rate equations for three
basic cases: {\bf subcritical} -- deleterious mutations favored ($b_->b_+$);
{\bf critical} -- no mutational bias ($b_+=b_-$); and {\bf supercritical} --
advantageous mutations favored ($b_+>b_-$).  In all three cases, the total
population density $N$ and the average age $A=N^{-1}\sum A_n$, with $A_n=\int
a\,C_n(a)\,da$, approach steady values.  These are determined by a balance
between the total birth rate $B$ and the death rate $\gamma N$ due to
overcrowding.  In the critical and supercritical cases, this leads to the
steady state behaviors for the total density and the average age,
\begin{equation}
\label{NA}
N={B\over \gamma}, \quad
A={1\over \gamma N}={1\over B}.
\end{equation}
The behavior in the subcritical case is more subtle, as we now discuss.

{\bf Subcritical Case.}  Here a steady state is reached whose properties are
found by setting ${dP_n\over dt}=0$ in Eq.~(\ref{pn}).  We solve this rate
equation by introducing the generating function $F(x)=\sum_{n\geq 1}
P_n\,x^{n-1}$ to transform the rate equation into the differential equation
\begin{equation}
\label{f2}
{F'\over F}={\gamma N-b+1-2b_-x\over b_{-}-(\gamma N-b)x
+b_+x^2}.
\end{equation}
Integrating Eq.~(\ref{f2}), subject to the obvious boundary condition
$F(1)=N$, gives a family of solutions which are parameterized by the total
population density $N$.  To extract a unique solution one has to invoke on
additional arguments.  First notice that $N$ lies within a finite range.  The
upper limit is found from the steady-state version of Eq.~(\ref{N}),
$(B-\gamma N)N= \Sigma n^{-1}P_n + b_-P_1 >0$, to give $\gamma N < B$.  The
lower limit is obtained from the physical requirement that all the $P_n$'s
are positive and therefore $F(x)$ is an increasing function of $x$.  From
Eq.~(\ref{f2}) this leads to the inequality $(\gamma N-b)^2\geq 4b_+b_-$.
Thus
\begin{equation}
\label{bounds}
b+2\sqrt{b_+b_-}\leq \gamma N<B.
\end{equation}
For any initial condition for the $P_n$ with a {\em finite} support in $n$,
only the minimal solution which satisfies the lower bound of
Eq.~(\ref{bounds}) is realized.  This selection is reminiscent of the
behavior in the Fisher-Kolmogorov equation and related reaction-diffusion
systems\cite{murray}.

To understand why the minimal solution is selected, consider the steady-state
asymptotic behavior of $P_n$ for $n\to\infty$.  In this limit, we may neglect
the $P_n/n$ term in Eq.~(\ref{pn}).  The resulting quasi-linear equation has
the solution $P_n=A_+\lambda_+^n+A_-\lambda_-^n$, with
$\lambda_\pm=\left[\gamma N-b\pm\sqrt{(\gamma N-b)^2-4b_+b_-}\right]/ 2b_-$,
and with $\lambda_\pm<1$.  Thus the steady-state fitness distribution decays
exponentially with $n$.  When the total population density attains the
minimal value $N_{\rm min}=(b+2\sqrt{b_+b_-})/\gamma$, $\lambda_+$ achieves
its minimum possible value $\lambda_+^{\rm min}=\sqrt{b_+/b_-}\equiv
\mu^{-1}$, where $\mu$ is the mutational bias.  Since $P_n\sim \lambda_+^n$,
the fitness distribution has the most rapid decay in $n$ for the minimal
solution.  This minimal solution appears to be the basin of attraction for
any initial condition with $P_n(0)$ decaying at least as fast as $\mu^{-n}$.
Conversely, an initial condition which decays as $\alpha^n$ with $\alpha$ in
the range $(\mu^{-1},\lambda_+^{\rm max}=1)$ should belong to the basin of
attraction of the solution where, from the steady-state version of
Eq.~(\ref{pn}) in the large-$n$ limit, the total population density is
$N=(b+b_-\alpha+b_+\alpha^{-1})/\gamma$.  We have verified this general
classification of solutions numerically\cite{hkr}.

\begin{figure}
\narrowtext
\epsfxsize=55mm\epsfysize=55mm
\hskip 0.35in\epsfbox{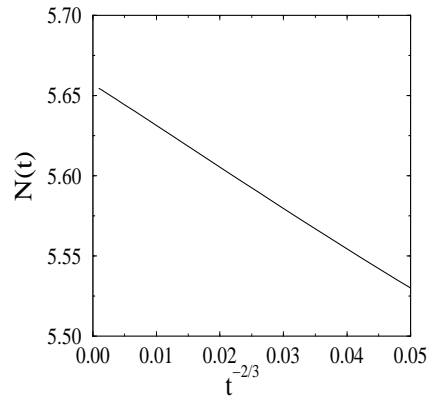}
\vskip 0.1in
\caption{$N(t)$ versus $t^{-2/3}$ in the subcritical case with $b=0$,
  $b_+=1$, $b_-=2$, and $\gamma=0.5$, and with the initial condition
  $P_n(t=0)=0.1$ for $1\leq n\leq 10$.    The asymptotic intercept
  with the $y$ axis gives the theoretically predicted value of $N_{\rm
    min}=4\sqrt{2}\approx 5.6568$.
\label{fig1}}
\end{figure}

Since the the steady state is approached exponentially in time for the
classical logistic equation, $\dot N=bN-\gamma N^2$, one might anticipate a
similar relaxation for our age-structured logistic equation (\ref{N}).
However, a numerical integration of the rate equations gives a power-law
relaxation of the total population density, $N_\infty-N(t) \sim t^{-2/3}$,
for a compact initial condition (Fig.~\ref{fig1}).  This is also verified by
an asymptotic analysis of the rate equations\cite{hkr}.  A similar relaxation
also occurs for the subpopulation densities with given fitness, $P_n(t)$.

For the relevant situation where the density $N$ takes the minimal value, we
integrate Eq.~(\ref{f2}) to give the generating function
\begin{equation}
\label{f3}
F(x)=N\left({\mu-1\over \mu-x}\right)^2 
\exp\left\{{x-1\over b_+(\mu-x)(\mu-1)}\right\}.
\end{equation}
One can formally determine the $P_n$ by expanding $F(x)$ in a Taylor series.
However, the asymptotic characteristics of the fitness distribution are more
easily determined directly from the generating function by using $\langle
n^k\rangle ={1\over N}\sum_{n=1}^\infty n^k\,P_n={1\over N}\, \left(x{d \over
    dx}\right)^k F(x)\Big|_{x=1}$.  Applying this to Eq.~(\ref{f3}), the
first two moments of the fitness distribution are
\begin{eqnarray}
\label{avf}
\langle n\rangle &=&{1\over b_+(\mu-1)^2}+{2\over \mu-1}+1,\nonumber \\
\sigma^2 &=&\langle n^2\rangle -\langle
n\rangle^2 ={\mu+1\over b_+(\mu-1)^3}+ {2\mu\over (\mu-1)^2}.
\end{eqnarray}

\begin{figure}
\narrowtext
\epsfxsize=55mm\epsfysize=55mm
\hskip 0.4in\epsfbox{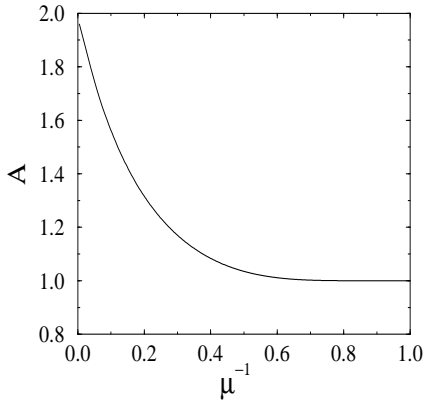}
\vskip 0.1in
\caption{Average age $A$ of the steady state population versus 
  $\mu^{-1}=\sqrt{b_+/b_-}$ with $b=0$, and with fixed total birth rate equal
  to one.  For $\mu^{-1}>1$, the average age $A=1$, while for $\mu^{-1}\to
  0$, $A\to 2$ \protect\cite{hkr}.
\label{fig2}}
\end{figure}

The average age of the population may be obtained by first solving
Eq.~(\ref{cn}) in the steady state to give
\begin{equation}
\label{cns}
C_n(a)=P_n\left(\gamma N+{1\over n}\right)\,
\exp\left[-\left(\gamma N+{1\over n}\right)a\right].
\end{equation}
The average age then is\cite{hkr}
\begin{eqnarray}
\label{age}
A&=&{1\over N}\sum_{n=1}^\infty \int_0^\infty a\,C_n(a)\,da,\nonumber\\
 &=&{1\over \gamma N}-{1\over N}\,{1\over (\gamma N)^2}
     \sum_{n=1}^\infty{P_n\over n+(\gamma N)^{-1}},\nonumber\\
 &=& {1\over \gamma N}-{1\over N}{1\over (\gamma N)^2}\,\int_0^1 
x^{1\over \gamma N}\,F(x)\,dx,
\end{eqnarray}
where the second line is obtained by using the expression for $C_n(a)$ from
Eq.~(\ref{cns}) and the last line follows by expressing the sum in terms of
an integral of the generating function.  The surprising feature that emerges
by numerical evaluation of this integral (Fig.~\ref{fig2}) is that the
average age {\em decreases} as the population gets fitter!

{\bf Supercritical Case.}  When $b_+>b_-$, the random walk in fitness space
defined by Eq.~(\ref{pn}) is biased away from the origin and a continuum
approach becomes appropriate in the long-time limit.  Treating $n$ as
continuous and Taylor expanding the master equation for small deviations
about $n$, gives the following convection-diffusion equation, supplemented by
birth/death terms, for the fitness distribution
\begin{equation}
\label{px}
\left({\partial \over \partial t}+V{\partial \over \partial n}\right)P
=\left(B-\gamma N-{1\over n}\right)P
+D{\partial^2 P\over \partial n^2}.
\end{equation}
The difference between the advantageous and deleterious mutation rates now
defines a bias velocity $V\equiv b_+-b_-$, and the average mutation rate
plays the role of a diffusion constant $D\equiv (b_++b_-)/2$.  Integrating
over all fitness values, the total population density obeys
\begin{equation}
\label{Nx}
{d N\over dt}=(B-\gamma N)N-\int_{0}^\infty {P(n,t)\over n}\,dn.
\end{equation}
Since the fitness distribution is sharply peaked at $\langle n\rangle=Vt$
(see below), the integral on the right-hand side approaches $N/Vt$.  By
setting ${dN\over dt}=0$ in the resulting equation, we conclude that $\gamma
N\to B-{1\over Vt}$.  This gives both the steady-state density, as well as
the rate of convergence to the steady state.

We now find the fitness distribution by substituting this asymptotics for
$N(t)$ into Eq.~(\ref{px}) to give
\begin{equation}
\label{px1}
\left({\partial \over \partial t}+V{\partial \over \partial n}\right)P
=\left({1\over Vt}-{1\over n}\right)P 
+D\,{\partial^2 P\over \partial n^2}.
\end{equation}
The birth/death term on the right hand side may be neglected, since $\langle
n\rangle=Vt$, and fluctuations about this average are of order $t^{1/2}$.
This approximation reduces Eq.~(\ref{px1}) into the classical
convection-diffusion equation, with solution
\begin{equation}
\label{Pappr}
P(n,t)={N\over \sqrt{4\pi Dt}}\,
\exp\left[-{(n-\langle n\rangle)^2\over 4Dt}\right].
\end{equation}
This gives a localized fitness distribution with average fitness growing
linearly in time, $\langle n\rangle=Vt$, and width growing diffusively,
$\sigma=\sqrt{2Dt}$.

To determine the age characteristics, notice that asymptotically, the $P_n$'s
change slowly with time, so that the time variable $t$ is {\it slow}.  On the
other hand, the age variable $a$ is {\em fast}.  Physically this reflects the
fact that during the lifetime of a typical individual the change in the age
characteristics of the population is small.  Thus in the first approximation,
we retain only the age derivative in Eq.~(\ref{cn}).  We also ignore the term
$C_n/n$, which is small near the peak of the asymptotic fitness distribution.
Solving the resulting master equation and using the boundary condition of
Eq.~(\ref{boundary}) we obtain
\begin{eqnarray}
\label{cnat}
C_n(a,t)&\simeq&P_n(t)\gamma N\,e^{-\gamma Na}\nonumber \\
&=&{\gamma N^2\over \sqrt{4\pi Dt}}\,
\exp\left[-\gamma Na-{(n-Vt)^2\over 4Dt}\right].
\end{eqnarray}
Summing over the fitness variable, the total age distribution $C(a,t)=\sum
C_n(a,t)$ is just a (stationary) Poisson, $C(a,t)=\gamma N^2\,e^{-\gamma
  Na}$, and the average age is $A=(\gamma N)^{-1}=B^{-1}$, in agreement with
Eq.~(\ref{NA}).

Let us compare this average age to that in the subcritical case; the latter
is given by Eq.~(\ref{age}) with $\gamma N=b+2\sqrt{b_+b_-}$.  To
provide a fair comparison (Fig.~\ref{fig2}), take the total birth rate $B$ to
be the same in both cases.  It can then be proved that the average age in the
supercritical case is always {\it smaller} than that in the subcritical
case\cite{hkr}.  Individuals in a population with preferential deleterious
mutations live longer than if advantageous mutations are favored!  The
continuous ``rat-race'' to increased fitness in the supercritical case does
not lead to an increase in the average life span.
 
{\bf Critical Case.}  With no mutational bias, the fitness still grows
indefinitely, but more slowly than in the supercritical system.  The equation
of motion for $P(n,t)$ is again given by Eq.~(\ref{px}), but with $V$ set
equal to zero and with $N(t)$ is still described by Eq.~(\ref{Nx}).  To
derive the scaling behaviors of $\langle n\rangle$ and the width of the
fitness distribution, we first use the fact that numerical integration
of Eq.~(\ref{px}) again gives a localized fitness distribution.  Thus we may
estimate the integral on the right-hand side of Eq.~(\ref{Nx}) as $N/\langle
n\rangle$.  This leads to $\gamma N\to B-{1\over \langle n\rangle}$.
Substituting this into Eq.~(\ref{px}) for $P(n,t)$ now yields
\begin{equation}
\label{p1}
{\partial P\over \partial t}
=\left({1\over \langle n\rangle}-{1\over n}\right)P 
+D{\partial^2 P\over \partial n^2}.
\end{equation}

To determine the long-time behavior of this equation, we exploit the fact
that the fitness distribution is peaked near $n\approx \langle n\rangle$.
This suggests changing variables from $(n,t)$ to the co-moving co-ordinates
$(y=n-\langle n\rangle,t)$.  Eq.~(\ref{p1}) then becomes
\begin{equation}
\label{p2}
{\partial P\over \partial t}-{d \langle n\rangle\over dt}\,
{\partial P\over \partial y}
={y\over \langle n\rangle^2}\,P-{y^2\over \langle n\rangle^3}\,P 
+D{\partial^2 P\over \partial y^2}.
\end{equation}
Let us first assume that the average fitness grows faster than diffusively,
that is, $\langle n\rangle\gg \sqrt{t}$.  With this assumption, the dominant
terms in Eq.~(\ref{p2}) are
\begin{equation}
\label{p3}
{d \langle n\rangle\over dt}\,{\partial P\over \partial y}
=-{y\over \langle n\rangle^2}\,P.
\end{equation}
These terms balance when $\langle n\rangle/(ty)\sim y/\langle n\rangle^2$.
Using this scaling in Eq.~(\ref{p2}) and then balancing the remaining
subdominant terms gives $y\sim \sqrt{t}$.  The combination of these results
then give $\langle n\rangle\sim t^{2/3}$.  This justifies our initial
assumption, $\langle n\rangle\gg \sqrt{t}$.  Finally, writing $\langle
n\rangle=(ut)^{2/3}$, simplifies Eq.~(\ref{p3}) to
\begin{equation}
\label{p4}
{\partial P\over \partial y}
=-{3y\over 2u^2 t}\,P,
\end{equation}
whose solution is the Gaussian of Eq.~(\ref{Pappr}), but with $\langle
n\rangle=(ut)^{2/3}$.  The value $u=\sqrt{3D}$ is determined by substituting
$\langle n\rangle=(ut)^{2/3}$ in Eq.~(\ref{p2}) and balancing the subdominant
terms.  To summarize, a Gaussian fitness distribution holds in both the
critical and supercritical cases with the fitness distribution peaked at
\begin{equation}
\label{fit}
\langle n\rangle=\cases{(3D)^{1/3}t^{2/3}  &critical case;\cr
                         Vt     &supercritical case.\cr}
\end{equation}

The age distribution in the critical case is obtained similarly to the
supercritical case.  The asymptotics of $C_n(a,t)$ is again given by a form
similar to Eq.~(\ref{cnat}), which gives $C(a,t)=\gamma N^2\,e^{-\gamma Na}$
after summing over $n$.  Hence the average age is $B^{-1}$, as in
Eq.~(\ref{NA}).

While our discussion is based on a population dynamics with an
age-independent mortality rate, this assumption does not substantially affect
our main results.  The crucial point is that old age is unattainable within
our model.  In the critical and supercritical cases, this is due to death by
increased competition among fit individuals, while in the subcritical case,
age is limited by the deleterious mutational bias.  Thus for a more realistic
mortality rate which increases with age, similar fitness and age dynamics to
those outlined here would still result\cite{hkr}.

In summary, in our population dynamics model, the average fitness grows
linearly in time when advantageous mutations are more likely and the fitness
approaches a steady value when deleterious mutations are favored.  In spite
of this fitness evolution, the average age of the population always reaches a
steady state.  Intriguingly, this average age is a decreasing function of the
average population fitness.  This paradoxical behavior arises because
competition becomes keener as the population becomes fitter.  

We gratefully acknowledge partial support from NSF grant DMR9632059 and ARO
grant DAAH04-96-1-0114.

\end{multicols}
\end{document}